\documentclass[twocolumn,showpacs,preprintnumbers,amsmath,amssymb]{revtex4}
\usepackage[T1]{fontenc}
\usepackage[latin1]{inputenc}

\usepackage{amsmath}
\usepackage{amssymb}
\usepackage{amsfonts}
\usepackage{graphicx}

\newcommand{\ket}[1]{\lvert{#1}\rangle}
\newcommand{\bra}[1]{\langle{#1}\rvert}
\newcommand{\braket}[2]{\langle#1\vert#2\rangle}

\begin{document}

\title{Efficiency of coherent state quantum cryptography in the presence of
  loss: \\Influence of realistic error correction}

\author{Matthias Heid}
\author{Norbert L\"utkenhaus}
\affiliation{Quantum Information Theory Group, Institut f\"ur theoretische
  Physik I and Max-Planck Research Group, Institute of Optics, Information and
  Photonics, Universit\"at Erlangen-N\"urnberg, Staudtstr. 7/B2, 91058
  Erlangen, Germany}
\date{\today}

\begin{abstract}
  We investigate the performance of a continuous variable (CV) quantum key
  distribution (QKD) scheme in a practical setting. More specifically, we take 
  non-ideal error reconciliation procedure into account. The quantum channel
  connecting the two honest parties is assumed to be lossy but
  noiseless. Secret key rates are given for the case that the measurement outcomes are postselected or a reverse reconciliation scheme
  is applied. The reverse reconciliation scheme loses its
  initial advantage in the practical setting. If one combines  postselection
  with reverse reconciliation however, much of this advantage can be
  recovered.
\end{abstract}

\maketitle
\section{Introduction}

Quantum key distribution (QKD) allows two parties, the sender Alice and the
receiver Bob, to share a key which is provably secure against any attack by an
eavesdropper (Eve), who may have superior computational and technological
power. Practical implementations of QKD use weak laser pulses, which can be 
easily controlled, or parametric down-conversion sources. These sources 
are used together with single-photon detectors to realize QKD. For a 
review see \cite{gisin02a}. Recently, it has been proposed to employ 
quadrature measurements of optical modes as detection devices, thereby 
introducing what is known as continuous variable (CV) QKD \cite{ralph00a,hillery00a,hirano03a,grosshans03a,silberhorn02a}. Some of 
these schemes use non-classical states, while others use coherent laser 
pulses. The aim is to exploit the high repetition
rates of homodyne detection  to overcome the limitation in detection rate that
is typical for single-photon counting at standard telecom wavelength although
recently other approaches using frequency conversion have been reported
\cite{diamanti05a, tanzilli05a}.

In the presence of loss, it seems to be
impossible at first sight to distill a secret key if the transmission falls
below 50\% (3dB loss) \cite{grosshans02a}. The reasoning is that Eve
can replace the lossy channel with an ideal one and use a beamsplitter to tap
off Alice's signals to simulate the losses. She then obtains the stronger
signals whenever the losses are bigger than 50\%. This apparent advantage of
Eve can be counteracted as it holds only on average, whether Alice and Bob are
closely or loosely correlated.  Alice and Bob can use postselection (PS) to
retain only those events where they are closely correlated and then have
some advantage over Eve. The distillation of a secret key rate is then
possible  for any transmittance of the quantum channel
\cite{silberhorn02b}. Another way to circumvent the 3dB loss limit is to use a
suitable one-way information reconciliation procedure. If one builds up the
key from Bob's measured data rather than from Alice's signals, Eve always has
less information about Bob's measurement result than Alice does. This
technique is known as reverse reconciliation (RR)
\cite{grosshans03a,grosshans03b} and leads to positive secret key rates
for an arbitrary loss of the quantum channel \cite{grosshans05a,navascues05a}.

There are several ways to extend this scenario to a more realistic one. First,
one could consider quantum channels that are not only lossy, but also impose
excess Gaussian noise on the quadrature distributions, as is
seen in experiments. For any excess channel noise $\delta$, as seen by Bob,
there exists a lower limit for  the tolerable single-photon transmittivity  $\eta$, which is given by \cite{namiki04a}
\begin{equation}
  \delta<2\eta \; .
\end{equation}
If the losses are higher, one can show that the data can be explained as
originating from an intercept resend attack. In this scenario, no secret key
can be distilled \cite{curty04a,curty05a}. In accordance with this bound, it has been shown that the key rate obtained  from PS-schemes decreases with increasing excess noise
\cite{namiki05a}.

Here we follow another direction to extend the work of
\cite{silberhorn02b,grosshans05a} and stick to the assumption of a lossy but
noiseless quantum channel. This scenario can be justified since detector
noise is the dominant contribution to the total noise seen in the experiment \cite{lorenz04a}. In a
trusted device scenario one can assume that Eve cannot exploit the noise of Bob's detectors. The remaining channel noise, which can leak information to
Eve, is typically less than one percent \cite{lorenz05suba} and can thus be neglected in a first
approximation. Similarly, in standard QKD with weak coherent pulses, the assumption of lossy but noiseless quantum channels together with detector dark counts which are inaccessible to Eve lead to a very good approximation to the rigorous secure key rate. In this scenario we investigate the
implications of the fact that any error
correction scheme in a real-world application cannot reach the fundamental
performance limit given by Shannon \cite{shannon48a}. The aim of this
article is to compare the performance of reverse reconciliation and
postselected schemes where the error correcting
carries a non-negligible overhead in the amount of necessary communication.

This paper is organized as follows: In the first section, we introduce the
investigated protocol and give the framework to which our efficiency
analysis applies. Next we calculate a lower bound on the secret key rate under
the assumption that the quantum channel between Alice and Bob is lossy but
noiseless and Eve is restricted to collective attacks. It turns out that
Eve's information about the key can be decomposed into effective binary
channels. Her information gain per use of such a binary channel is computed
in section III A for the protocol using direct reconciliation (DR) and for the reverse reconciled protocol (RR) in section III B. Afterwards, we calculate the total secret key rates for the various protocols and include the possibility to use postselection
(PS). We then proceed by including inefficient but trusted
detectors on Bob's side in our analysis in section IV. Details about the
numerical optimization of the secret key rates are given in section V.  In the last section we conclude and discuss our results.

\section{Protocol and beamsplitter attack}

We consider the situation where Alice sends pure signal states
$\ket{\phi_{i}}$ through the quantum channel to Bob, who can verify, for
example by performing tomographic complete measurements, that he indeed always
receives pure conditional states as expected in the absence of channel noise $\ket{\Psi_{i}}$ \footnote{For a full
  security proof one would have to make the point more precise. However, this
  is not the aim of this letter.}. It follows that
the bipartite state of Eve and Bob has to be a product state. In the scenario
of collective attacks, the action of the channel can be modeled by a unitary
coupling to an ancilla system accessible to Eve,  which is prepared in some
standard state $\ket{\epsilon^{0}}$. The absence of noise allows to fully characterize Eve's attack.  We have 
\begin{eqnarray*}
U\ket{\phi_{i}}\ket{\epsilon^{0}}=\ket{\Psi_{i}}\ket{\epsilon_{i}}.
\end{eqnarray*}
Since $U$ is unitary, we have
\begin{eqnarray*}
\braket{\phi_{i}}{\phi_{j}}=\braket{\Psi_{i}}{\Psi_{j}}\braket{\epsilon_{i}}{\epsilon_{j}},
\end{eqnarray*}
where $\braket{\phi_{i}}{\phi_{j}}$ is given by the state preparation and
$\braket{\Psi_{i}}{\Psi_{j}}$ is fixed by Bob's observation. Therefore the
overlaps $\braket{\epsilon_{i}}{\epsilon_{j}}$, which contain all of Eve's information about the signals, are fixed.

In our protocol, Alice encodes her bit-value into the modulation of two
coherent states according to 
\begin{eqnarray*}
  \ket{\underline{0}}&=&\ket{\alpha}\\
  \ket{\underline{1}}&=&\ket{-\alpha},
\end{eqnarray*}
with the coherent amplitude $\alpha$ chosen to be real without loss of
generality. The states are sent with equal a priori probabilities
$\mathrm{p}_{0}=\mathrm{p}_{1}=\frac{1}{2}$. 
Bob performs a heterodyne measurement on the received states, which is
mathematically equivalent to a projection onto a coherent state
$\ket{\beta}=\ket{\beta_{x}+i \beta_{y}}$. From that he is able to conclude
that he indeed received pure states. That in turn fixes Eve's knowledge about
the signal states.  Furthermore, if Bob receives attenuated coherent states
$\ket{\pm\sqrt{\eta}\alpha}$, the states that Eve holds have to be unitarily
equivalent to those obtainable by the beamsplitter attack. The input states
are transformed according to
\begin{equation}\label{BSattack}
\ket{\pm \alpha} \rightarrow \ket{\pm \sqrt{\eta}\alpha}_{B}\otimes \ket{\pm \sqrt{1-\eta}\alpha}_{E}
\end{equation}
in this attack. After measuring the signals, Bob assigns the
bit-value 0 (1), if $\beta_{x}$ is positive (negative) and then publicly
announces $\beta_{y}$ and the modulus of $\beta_{x}$ of the measured
$\beta$. As we will see, this announcement will enable us to decompose the protocol into effective binary information channels.

\section{Lower bound on secret key rate}
After the exchange of quantum signals between Alice and Bob is
complete, they  proceed with a classical post-processing phase in which
they correct for errors in their bit-strings and cut out Eve's knowledge about
the key (privacy amplification)\cite{bennett95a}. To do so, they need to transmit information though an authenticated
but otherwise insecure classical channel. Eve may listen and use any
information exchanged over the public channel to optimize her attack.

Alice and Bob may use the classical channel only in one direction for error correction. This will result
in two non-equivalent ways of distilling a secret key from their shared
classical data. Communication from Alice to Bob is common in QKD and we will
refer to it as direct reconciliation (DR), whereas communication in the
opposite direction is called reverse reconciliation (RR) \cite{grosshans03a}. The secret key is
built from the data that the sender in the classical communication step
holds. In any case, Eve's knowledge is summarized in quantum states
$\rho_{i}$ conditioned on bit-values held by the person who transmits the
error correction information. Her knowledge about the data can be quantified by the Holevo quantity $\chi$ \cite{holevo73b}, given by
\begin{eqnarray}\label{Holevo}
\chi &=& S(\overline{\rho})-\sum_{i=0}^{1}\mathrm{p}_{i} S(\rho_{i})\\
\overline{\rho}&=& \sum_{i=0}^{1} \mathrm{p}_{i}\rho_{i}\nonumber,
\end{eqnarray}
which includes Eve being allowed to measure out her ancillas collectively.
 It turns out  \cite{devetak05a} that the secret
key rate $G$ in this collective attack scenario will then be bounded from below by
\begin{equation}\label{Dev/Win}
G \geq \mathrm{I}_{A:B}-\chi \; .
\end{equation}
Note that we have replaced the Holevo quantity between Alice and Bob in theorem 1 of \cite{devetak05a} by the classical mutual
Information $\mathrm{I}_{A:B}$, since we are investigating a practical QKD
scheme with our specified measurement setup.

Next, we will evaluate $\mathrm{I}_{A:B}$ and $\chi$ for different protocols and noiseless detectors.

\subsection{Mutual information between Alice and Bob}

After the quantum states are distributed and measured, Alice and Bob
share classical correlated bit-strings. The mutual information
$\mathrm{I}_{A:B}$ between the two honest parties is determined by the
conditional probabilities that Bob projects onto $\beta$. These are given by
\begin{eqnarray}\label{condprob}
\mathrm{p}(\beta|0)&=&\frac{1}{\pi}\mathrm{e}^{-\left(\left(\beta_{x}-\sqrt{\eta}\alpha\right)^2+\beta_{y}^2\right)}\\
\mathrm{p}(\beta|1)&=&\frac{1}{\pi}\mathrm{e}^{-\left(\left(\beta_{x}+\sqrt{\eta}\alpha\right)^2+\beta_{y}^2\right)}\nonumber.
\end{eqnarray}
Since Eve is only performing an individual coupling of the signals to her
ancilla systems, Bob's measurement outcomes $\beta$
for different signals are independent. The mutual information
$\mathrm{I}_{A:B}$ between Alice and Bob cannot depend on the value of
$\beta_{y}$, since the agreement on a bit-value does not
depend on it. Furthermore, the total probability that Bob obtains the
measurement outcome $\beta_{x}$ is given by
\begin{eqnarray}\label{marginal}
 \mathrm{p}(\beta_{x})&=&\int_{-\infty}^{\infty}\mathrm{d}\beta_{y}
 \; \mathrm{p}(\beta)=\\&=&\frac{1}{2\sqrt{\pi}}\left(\mathrm{e}^{-\left(\beta_{x}+\sqrt{\eta}\alpha\right)^2}+\mathrm{e}^{-\left(\beta_{x}-\sqrt{\eta}\alpha\right)^2}\right)\nonumber.
\end{eqnarray}
From this we see that the two outcomes $\pm \beta_{x}$ occur with the same
probability. It follows that the announcement of $\beta_y$ and $|\beta_{x}|$
defines an effective binary information channel. The probability $e^{+}$ that Bob
assigns the wrong bit-value for a given positive value of $\beta_{x}>0$ is given by
\begin{eqnarray*}
e^{+}=
\frac{\mathrm{p}\left(\beta_{x}|1\right)}{\mathrm{p}\left(\beta_{x}|0\right)+\mathrm{p}\left(\beta_{x}|1\right)} ,
\end{eqnarray*}
whereas the corresponding error probability $e^{-}$ for negative outcomes
$\beta_{x}<0$ is given by
\begin{eqnarray*}
e^{-}=
\frac{\mathrm{p}\left(\beta_{x}|0\right)}{\mathrm{p}\left(\beta_{x}|0\right)+\mathrm{p}\left(\beta_{x}|1\right)} .
\end{eqnarray*}
From Eqn. (\ref{condprob}) it follows that 
\begin{eqnarray}\label{erate}
e^{+}=e^{-}=e=\frac{1}{1+\mathrm{e}^{4\sqrt{\eta}\alpha|\beta_{x}|}},
\end{eqnarray}
so that the effective information channels are symmetric in the error rate.
Each information channel contributes an
amount of $1-\mathrm{H}^{\mathrm{bin}}(e)$ to the mutual information, $\mathrm{I}_{A:B}$,
where  $\mathrm{H}^{\mathrm{bin}}(e)$ is the entropy of the
binary symmetric channel,
\begin{eqnarray*}
 \mathrm{H}^{\mathrm{bin}}(e)=-e\;\mathrm{log}_{2}(e)-(1-e)\;\mathrm{log}_{2}(1-e).
\end{eqnarray*}
The probability that an effective information channel is being used, is given by
\begin{equation}\label{probchannel}
  \mathrm{p}_{c}(\beta_{x})=2\mathrm{p}(\beta_{x}).
\end{equation}
 For the total transmission  we find
\begin{equation}
\label{totalinfo}
\mathrm{I}_{A:B}= \int_{0}^{\infty} \mathrm{d}\beta_{x}
 \;\mathrm{p}_{c}(\beta_{x})\left[1-\mathrm{H}^{\mathrm{bin}}\left(e\right) \right] \;.
\end{equation}
In order to calculate the secret key rate $G$ according to Eqn. 
(\ref{Dev/Win}), we proceed by bounding Eve's knowledge $\chi$ decomposed in
the effective binary information channels.

\subsection{Direct reconciliation}

Usually in QKD the DR case is considered where the secret key is determined by Alice's data. This means that Alice sends Bob error correction information in the
information reconciliation step of the protocol. After Alice and Bob have
corrected their bit-strings, Eve can make use of the information transmitted
over the public channel to optimize her measurements on her ancilla
systems. 

The quantum states in Eve's hand, conditioned on Alice's data, are given by (\ref{BSattack}) as
\begin{equation}\label{condDR}
\ket{\epsilon_{i}}=\ket{\pm\sqrt{1-\eta}\alpha}\; .
\end{equation}
These states are pure, so that we have 
$\chi^{DR} = S(\overline{\rho})$. What remains to be calculated are the eigenvalues of 
\begin{equation*}
\overline{\rho}=\frac{1}{2}\left(\ket{\epsilon_{0}}\bra{\epsilon_{0}}+\ket{\epsilon_{1}}\bra{\epsilon_{1}}\right).
\end{equation*}
The symmetry allows us to write the states $\ket{\epsilon_{i}}$ as
\begin{eqnarray}\label{sym}
\ket{\epsilon_{0}}&=&c_{0}\ket{\Phi_{0}}+c_{1}\ket{\Phi_{1}}\\
\ket{\epsilon_{1}}&=&c_{0}\ket{\Phi_{0}}-c_{1}\ket{\Phi_{1}}\nonumber,
\end{eqnarray}
where the $\ket{\Phi_{i}}$ are orthonormal states. A short calculation shows
that $\overline{\rho}$ is already diagonal in this basis with eigenvalues
$\left|c_{i}\right|^2$, so that the Holevo quantity is given by
\begin{eqnarray}\label{entropyrho}
&&\chi^{DR}=S(\overline{\rho})=-\sum_{i=0}^{1} \left|c_{i}\right|^2 \mathrm{log} \left(\left|c_{i}\right|^2\right).
\end{eqnarray}
The normalization of $\rho$
\begin{eqnarray*}
\left|c_{0}\right|^2+\left|c_{1}\right|^2=1
\end{eqnarray*}
and the overlap
\begin{eqnarray*}
\left|c_{0}\right|^2-\left|c_{1}\right|^2=\braket{\epsilon_{0}}{\epsilon_{1}}
\end{eqnarray*}
give the  expression for the coefficients
\begin{eqnarray}
\left|c_{0}\right|^2&=&\frac{1}{2}\left(1+\braket{\epsilon_{0}}{\epsilon_{1}}\right)\label{c0c1def}\\
\left|c_{1}\right|^2&=&\frac{1}{2}\left(1-\braket{\epsilon_{0}}{\epsilon_{1}}\right).\nonumber
\end{eqnarray}
The overlap of the two coherent states (\ref{condDR}) is
\begin{eqnarray}\label{eveoverlap}
\braket{\epsilon_{0}}{\epsilon_{1}}=\mathrm{e}^{-2(1-\eta)\alpha^2},
\end{eqnarray}
so that the Holevo quantity can be directly computed.
Since Eve's quantum states are independent of Bob's measurement outcomes (see
formula (\ref{eveoverlap})), the value of the Holevo quantity for each
effective channel,  $\chi^{DR}(\beta_{x})$, does not depend on $\beta$. Therefore, we find that the total Holevo quantity for all effective channels becomes 
\begin{equation}
\label{DRchannel}
\chi^{DR}=\chi^{DR}(\beta_{x})\;.
\end{equation}

\subsection{Reverse reconciliation}

In a reverse reconciliation scheme \cite{grosshans03a}, Bob sends error correction information to
Alice.  Therefore Eve is interested in obtaining information about Bob's measurement results rather than  Alice's signals. After Bob's announcement of
$|\beta_{x}|$ and $\beta_{y}$, Eve knows which information channel is used and, depending on Bob's measurement result,that she either holds the state
\begin{eqnarray*}
 \rho_{+}=\left(1-e\right)\ket{\epsilon_{0}}\bra{\epsilon_{0}}+e\ket{\epsilon_{1}}\bra{\epsilon_{1}}
\end{eqnarray*} or
\begin{eqnarray*}
 \rho_{-}=e\ket{\epsilon_{0}}\bra{\epsilon_{0}}+\left(1-e\right)\ket{\epsilon_{1}}\bra{\epsilon_{1}}
\end{eqnarray*} 
in her ancilla system. Here  $e$ is the error rate of the effective binary
information channel (\ref{erate}). In order
to obtain the correct bit-value, she needs to find out the sign of $\beta_{x}$. Both signs of the measurement outcome occur with the same
probability $\mathrm{p}(\beta_{x})$ of Eqn. (\ref{marginal}), so that the
probability of the effective information channel being used is again
$\mathrm{p}_{c}=2 \mathrm{p}(\beta_{x})$.  The Holevo quantity (\ref{Holevo}) in the case of reverse reconciliation reads
\begin{eqnarray*}
\chi^{RR}(\beta_{x})&=&S(\overline{\rho})-\mathrm{p}_{+}S(\rho_{+})-\mathrm{p}_{-}S(\rho_{-})\\
&=&S(\overline{\rho})-\frac{1}{2}\left[S(\rho_{+})+S(\rho_{-})\right],
\end{eqnarray*}
since the probability that Bob gets a positive (negative) outcome for a given
binary information channel, is simply $\mathrm{p}_{\pm}=\frac{1}{2}$ (see
Eqn. \ref{marginal}).
The entropy of $\overline{\rho}$ has already been calculated and is given by
Eqns. (\ref{entropyrho}) and (\ref{c0c1def}).

Furthermore there exists a unitary operation $U$ (a phase-shift of $\pi$) with
\begin{eqnarray*}
\rho_{+}=U\rho_{-}U^{\dagger},
\end{eqnarray*}
so that $S(\rho_{+})=S(\rho_{-})$.
The basis states $\ket{\Phi_{i}}$ of Eqn. (\ref{sym}), which are adapted to
the symmetry, can again be used to calculate $S(\rho_{+})$. In this basis
$\rho_{+}$ reads
\begin{eqnarray*}
\rho_{+}=\left(
\begin{array}{cc}\left|c_{0}\right|^2& \left(1-2e\right)c_{1}^{*}c_{0}\\
  \left(1-2e\right)c_{0}^{*}c_{1}&\left|c_{1}\right|^2
\end{array}
\right).
\end{eqnarray*}
One can show with the help of Eqn. (\ref{c0c1def}) that the eigenvalues
$\lambda_{1,2}$ of this matrix are of the form
\begin{eqnarray}\label{eig}
  \lambda_{1,2}=\frac{1}{2}\left(1\pm\sqrt{1+4e\left(e-1\right)(1-|\braket{\epsilon_{0}}{\epsilon_{1}}|^2)}\right),
\end{eqnarray}
so that we have explicitly given
\begin{eqnarray}\label{entrrho+}
  S(\rho_{+})=\sum_{i=1}^{2}-\lambda_{i}\mathrm{log}\lambda_{i}.
\end{eqnarray}
With the help of Eqns. (\ref{entropyrho}),(\ref{eveoverlap}), (\ref{eig})
and (\ref{entrrho+}), the Holevo quantity per information channel can now be
explicitly evaluated via
\begin{eqnarray}\label{HolevoRR2}
\chi^{RR}(\beta_{x}) =S(\rho)-S(\rho_{+}).
\end{eqnarray}
This quantity varies for the different effective channels. 

\subsection{Postselection}

We have calculated an upper bound $\chi$ of Eve's information about the
key for a given information channel for DR  and for RR. Also, we have an expression for the total mutual information shared between the two parties, Eqn. (\ref{totalinfo}).  The
total achievable key rate per signal, as given by  Eqn. (\ref{Dev/Win}),  can then be written as
\begin{eqnarray}\label{keyideal}
 \mathrm{G}&=&\int_{0}^{\infty} \mathrm{d}\beta_{x}
 \;\mathrm{p}_{c}(\beta_{x})\underbrace{\left[1-\mathrm{H}^{\mathrm{bin}}\left(e\right)-\chi(\beta_{x})
\right]}_{=:\Delta\mathrm{I}^{\mathrm{ideal}}(\beta_{x}) } \; ,
\end{eqnarray}
where $\chi(\beta_x)$ is given by Eqns. (\ref{entropyrho}) and (\ref{DRchannel}) for DR and by formula
(\ref{HolevoRR2}) for RR. Here the sum runs over all possible information
channels. In principle one can improve the performance of the protocols by
dismissing channels where $\Delta\mathrm{I}^{\mathrm{ideal}}(\beta_{x})<0$,
since Eve learns more on average about the
signals than Bob for those values of $\beta_{x}$. This procedure is called postselection. 

The key rate (\ref{keyideal}) refers to the case where a perfect
error correction procedure is used (Shannon limit). Ideally, in order to
correct a bit-string of large length $n$, one has to exchange asymptotically $n \mathrm{H}^{\mathrm{bin}}(e)$ bits over
the public channel. This information has to be hidden from Eve, which can be
done in principle by using a one-time-pad of exactly the same length. In the
end, each use of an information channel with error rate $e$ costs
$\mathrm{H}^{\mathrm{bin}}(e)$ secret bits to encrypt the necessary error
correction information
\footnote{Alternatively, one can think of Eve simply learning
  $\mathrm{H}^{\mathrm{bin}}(e)$ bits per exchanged signal during error
  correction. This amount of information has then to be cut out in the privacy
  amplification steps. The final formulas do not change,
  however \cite{cachin97a}.}. 

It turns out that all effective information channels yield a positive
  contribution $\Delta\mathrm{I}^{\mathrm{ideal}}_{RR}(\beta_{x}) \geq 0$ for the
  RR protocols in this idealized setting. This is in agreement with the result
  found in \cite{grosshans05a}. Therefore, it is possible to distill
  a secret key for any transmission of the quantum channel, and the
  performance of the RR protocol cannot be improved further by using postselection.

Practical codes that work exactly at the Shannon limit are not
known. Efficient codes work close to that
limit, so in practice one has to reveal more
information to correct one bit, i.e.  $f(e)\mathrm{H}^{\mathrm{bin}}(e)$,
where $f(e)$ represents the efficiency of the used protocol ($f(e)\geq1$). The
coefficient $f(e)$, which  determines the overhead one has to pay for realistic error correction, depends in all practical schemes on the error
rate $e$. To see how the key rate $G$ scales when an error correction scheme
with efficiency $f(e)$ is used, we can rewrite formula (\ref{keyideal}) as
\begin{eqnarray}\label{keyprac}
 \mathrm{G}&=&\int_{0}^{\infty} \mathrm{d}\beta_{x}
 \;\mathrm{p}_c(\beta_{x})\left[\underbrace{1-f(e)\mathrm{H}^{\mathrm{bin}}(e)-\chi(\beta_{x})}_{=:\Delta\mathrm{I}^{\mathrm{prac}}(\beta_{x})} \right] .
\end{eqnarray}
The quantity $\chi(\beta_{x})$ is again given by Eqns. (\ref{entropyrho}) and (\ref{DRchannel}) in the DR and
by (\ref{HolevoRR2}) in the RR case respectively. Postselection can again be
applied once Alice and Bob know the efficiency $f(e)$ of their error correction
procedure approximately. Thus postselection now is relevant not only for DR, but also for the RR scenario.

\section{Detector noise}

By now we have obtained secret key rates while neglecting any kind of
noise. While experiments show that the channel noise is low, the noise of the
detector is not negligible, but of the order of 0.1 shot noise units. In a
trusted device scenario we assume that Eve cannot manipulate the detector
noise to leak information about the signals to her. On the other hand, this
noise increases the cost of error correction. It is interesting to see how
this affects the key rate. We define the excess noise $\delta$ imposed by the detector by
\begin{equation}\label{excess}
\delta=\frac{\Delta^2_{\mathrm{obs}}\beta_{x}}{\Delta^2_{\mathrm{SNL}}\beta_{x}}-1,
\end{equation}
where $\Delta^2_{\mathrm{obs}}\beta_{x}$ is the observed variance of
  $\beta_{x}$ seen in experiments and $\Delta^2_{\mathrm{SNL}}\beta_{x}$
  is the shot noise limited variance of $\beta_{x}$.
The probability that Bob obtains the measurement outcome $\beta_{x}$ is then
given by
\begin{eqnarray}\label{modprob}
 &&\mathrm{p^{\mathrm{Det}}}(\beta_{x})=\\&&\frac{1}{2\sqrt{\pi(1+\delta)}}\left(\mathrm{e}^{\frac{-\left(\beta_{x}+\sqrt{\eta}\alpha\right)^2}{1+\delta}}+\mathrm{e}^{\frac{-\left(\beta_{x}-\sqrt{\eta}\alpha\right)^2}{1+\delta}}\right)\nonumber \;,
\end{eqnarray}
including detector noise. This leads to a modified error rate of the efficient binary channels, given by
\begin{equation}\label{moderate}
e^{\mathrm{Det}}=\frac{1}{1+\mathrm{e}^{\frac{4\sqrt{\eta}\alpha|\beta_{x}|}{1+\delta}}}.
\end{equation}
The secret key rate can then be calculated as described in the previous
sections for any value of the excess noise $\delta$, one has only to use the
modified expressions for Bob's probability distribution (\ref{modprob}) and for
the error rate (\ref{moderate}) instead of Eqns. (\ref{marginal}) and
(\ref{erate}) in all preceding formulas. This will lead to a decrease of the
mutual information $\mathrm{I}_{AB}$ between Alice and Bob with growing excess
noise $\delta$ to account for the higher cost of error correction. While the
cost of privacy amplification $\chi^{DR}$ per use of an effective binary
information channel in the DR protocol (\ref{DRchannel}) is unaffected by the
detector noise, the corresponding quantity $\chi^{RR}$ (\ref{HolevoRR2}) for
the RR protocol decreases with increasing detector noise. This effect
originates from the fact that Eve is more uncertain about Bob's measurement
outcomes if he uses inefficient detectors. 

\section{Numerical procedure for calculating the key rate}

Now we have everything at hand to compute $G$. For this, we resort to
numerical calculations. We assume that our
error correction can work as efficiently as the bidirectional protocol Cascade \cite{brassard93a}. To be precise, we use a linear fit
of the efficiency of Cascade (see table \ref{effcascade}) for the function $f(e)$ in our numerical
optimization. 
  \begin{table}
    \begin{center}
      \begin{tabular}{c|c}
	$e$ & $f(e)$\\
	\hline \hline
	0.01 & 1.16\\
	0.05 & 1.16\\
	0.1 & 1.22\\
	0.15 & 1.32\\
      \end{tabular}
      \caption{\label{effcascade} Efficiency of Cascade \cite{brassard93a} for different values of
	the error rate $e$}
    \end{center}
  \end{table}
Formally, one-way communication would be needed to justify the use of the
Devetak-Winter bound (\ref{Dev/Win}). However in the worst case scenario when
using two-way communication, Eve learns all positions where Bob assigned the wrong bit-value. Therefore the Devetak-Winter bound could still be applied if
one additionally announces Bob's error positions. But since Eve's knowledge
about the key (\ref{entropyrho}) in the DR setting does not depend on Bob's measurement
outcome $\beta_{x}$, the resulting key rate would remain unchanged. Therefore two-way error correction methods can be applied directly in
connection with DR methods in the lossy channel.  In RR, however, 
strict one-way communication is essential. Since we are here interested in
efficiency considerations only, we ignore the problem of finding a practical
one-way protocol that can be as efficient as Cascade and simply assume we have such a protocol.  Still, we have to keep
in mind that the rate for DR can be implemented directly with known
protocols, whereas the RR rate requires the usage of efficient one-way error
correction protocols.

For given transmission $\eta$  of the quantum channel and excess noise
$\delta$ imposed by Bob's detector we 
\begin{itemize}
  \item choose an amplitude  $\alpha$ of Alice's coherent signal states,
  \item calculate $\Delta \mathrm{I}^{\mathrm{prac}}(\beta_{x})$ for all information channels
  $\beta_{x}$ taking realistic error correction into account,
  \item discard all channels where $\Delta \mathrm{I}^{\mathrm{prac}}(\beta_{x})<0$ in
  postselection protocols,
  \item integrate over the remaining channels to obtain the key rate $G(\eta,\alpha)$,
  \item start over such as to  optimize over $\alpha$ for a given transmission $\eta$.
\end{itemize}

\section{Numerical evaluation and discussion}

\begin{figure}
  \begin{center}
  \includegraphics[width=0.48\textwidth]{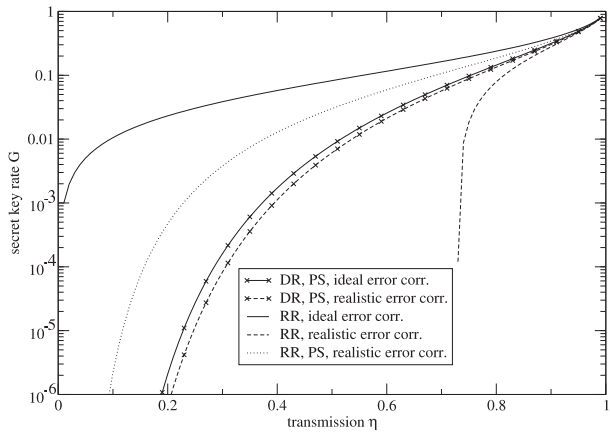}
  \end{center}
  \caption{\label{figure} Comparison of the secret key rate $G$ versus
  transmission $\eta$ for ideal (solid lines) and realistic error correction
  (dashed lines) for a postselected DR protocol and a RR protocol. The dotted
  line represents a postselected RR protocol with realistic error correction.}
\end{figure}

Let us first neglect any detector noise on Bob's side. Fig. \ref{figure}
summarizes our numerical results for the RR protocol and for the postselected
DR scheme for the case that Bob's detectors are noiseless.
If ideal error correction is assumed, the RR scheme clearly performs 
better than the postselected DR scheme. In this setting, an additional
postselection step in the RR protocol cannot improve the performance, since
all information channels yield a positive advantage $\Delta
\mathrm{I}^{\mathrm{ideal}}_{RR}(\beta_{x})\geq 0$ for Alice and Bob. The key rate for a
non-postselected DR scheme is not shown, since it would be limited by 50\%
losses and is therefore not of interest for practical QKD.

The key rate decreases significantly in the non-postselected RR protocols if one does not
assume an ideal error correction. This is however not a problem of RR itself,
but due to the fact that in the simple approach all bits have to be corrected. Since one has to
shrink the key by $f(e)\mathrm{H}^{\mathrm{bin}}(e)$ bits of information to
correct a bit coming from a channel with error rate $e$ in a realistic
scenario, it follows that the usage of information channels with high error
rate $e$ effectively shrinks the key. Using all information channels would completely
negate the advantage of non-postselected RR versus postselected DR in a realistic scenario. As a remedy, we propose to introduce
postselection of effective binary channels in the RR protocol as well. This
combination of the RR idea and postselection can help to give good performance with realistic protocols. The dotted curve in
Fig. \ref{figure} represents our numerical results for a postselected RR
protocol.

\begin{figure}
\includegraphics[width=0.48\textwidth]{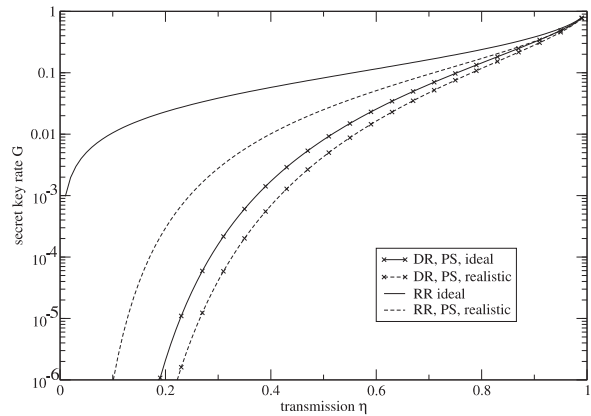}
\caption{\label{detector}  Influence of detector noise. Secret key rates $G$ versus transmission $\eta$ for ideal (solid lines) and realistic error correction
  (dashed lines) for a postselected DR protocol and a postselected RR protocol
  are shown. The dashed lines include an excess noise of $\delta=0.1$.}
\end{figure}

Our results for imperfect detectors are summarized in Fig.
\ref{detector}. Since it is necessary to postselect the data in the RR case
even in absence of detector noise, we omit the curves for the
non-postselected RR protocol. We find that all investigated postselected
protocols are robust against typical values of the detector excess noise
$\delta$. The dashed lines in Fig. \ref{detector} include a detector noise
of $\delta=0.1$ and do not differ significantly from the corresponding curves
in Fig. \ref{figure} which include an inefficient error
correction protocol but neglect noisy detectors. For the RR protocol, the
effect of detector noise is almost negligible. As mentioned before, this is due
to the fact that the cost of privacy amplification $\chi^{RR}$ decreases in
the RR scenario. This can partially compensate for the higher cost of error correction. 

In conclusion we find that it is important to take the influence of
inefficient error correction into account in evaluating QKD protocols. We propose to combine postselection
with reverse reconciliation to deal with losses in realistic continuous
variable QKD. It should be emphasized that efficient one-way error
reconciliation procedures are essential to make this approach work. As a
fall-back position, we can use the DR scheme with PS for which protocols are
available already today. 

We thank F. Grosshans for helpful discussions. This work has been supported by
the network of competence QIP of the state of Bavaria (A8), the EU-IST network
SECOQC and the German Research Council (DFG) under the Emmy-Noether program.

\end{document}